\documentstyle[12pt]{article}
\ifx\optionkeymacros\undefined\else \fi

\catcode`\Œ=\active\defŒ{{\aa}}       
\catcode`\º=\active\defº{\int}        
\catcode`\=\active\def{\c c}        
\catcode`\¶=\active\def¶{\partial}    
\catcode`\Ä=\active\defÄ{\oint}       
\catcode`\Æ=\active\defÆ{\triangle}   
\catcode`\Â=\active\defÂ{\neg}        
\catcode`\µ=\active\defµ{\mu}         
\catcode`\¿=\active\def¿{{\o}}        
\catcode`\¹=\active\def¹{\pi}         
\catcode`\Ï=\active\defÏ{{\oe}}       
\catcode`\§=\active\def§{{\ss}}       
\catcode`\ =\active\def {\dagger}     
\catcode`\Ã=\active\defÃ{\sqrt}       
\catcode`\·=\active\def·{\Sigma}      
\catcode`\Å=\active\defÅ{\approx}     
\catcode`\½=\active\def½{\Omega}      
\catcode`\£=\active\def£{{\it\$}}     
\catcode`\°=\active\def°{\infty}      
\catcode`\¤=\active\def¤{{\S}}        
\catcode`\¦=\active\def¦{{\P}}        
\catcode`\¥=\active\def¥{\bullet}     
\catcode`\»=\active\def»{\leavevmode\raise.585ex\hbox{\b a}}      
\catcode`\¼=\active\def¼{\leavevmode\raise.6ex\hbox{\b o}}        
\catcode`\­=\active\def­{\not=}       
\catcode`\²=\active\def²{\leq}        
\catcode`\³=\active\def³{\geq}        
\catcode`\Ö=\active\defÖ{\div}        
\catcode`\É=\active\defÉ{{\dots}}     
\catcode`\¾=\active\def¾{{\ae}}       
\catcode`\Ç=\active\defÇ{\ll}         
\catcode`\Ò=\active\defÒ{``}          
\catcode`\Á=\active\defÁ{!`}          
\catcode`\¢=\active\def¢{\rlap/c}     
\catcode`\Ô=\active\defÔ{`}           
\catcode`\Õ=\active\defÕ{'}           

\catcode`\=\active\def{{\AA}}       
\catcode`\'=\active\def'{\c C}        
\catcode`\¯=\active\def¯{{\O}}        
\catcode`\¸=\active\def¸{\Pi}         
\catcode`\Î=\active\defÎ{{\OE}}       
\catcode`\®=\active\def®{{\AE}}       
\catcode`\×=\active\def×{\diamond}    
\catcode`\¡=\active\def¡{\accent'27}  
\catcode`\Ó=\active\defÓ{''}          
\catcode`\±=\active\def±{\pm}         
\catcode`\È=\active\defÈ{\gg}         
\catcode`\À=\active\defÀ{?`}          
\catcode`\Ð=\active\defÐ{--}          
\catcode`\Ñ=\active\defÑ{---}         

\catcode`\Š=\active\defŠ{\"a}        
\catcode`\'=\active\def'{\"e}        
\catcode`\•=\active\def•{\"{\i}}     
\catcode`\š=\active\defš{\"o}        
\catcode`\Ÿ=\active\defŸ{\"u}        
\catcode`\Ø=\active\defØ{\"y}        
\catcode`\€=\active\def€{\"A}        
\catcode`\…=\active\def…{\"O}        
\catcode`\†=\active\def†{\"U}        
\catcode`\‡=\active\def‡{\'a}        
\catcode`\Ž=\active\defŽ{\'e}        
\catcode`\'=\active\def'{\'{\i}}     
\catcode`\—=\active\def—{\'o}        
\catcode`\œ=\active\defœ{\'u}        
\catcode`\ƒ=\active\defƒ{\'E}        
\catcode`\ˆ=\active\defˆ{\`a}        
\catcode`\=\active\def{\`e}        
\catcode`\"=\active\def"{\`{\i}}     
\catcode`\˜=\active\def˜{\`o}        
\catcode`\=\active\def{\`u}        
\catcode`\Ë=\active\defË{\`A}        
\catcode`\‹=\active\def‹{\~a}        
\catcode`\–=\active\def–{\~n}        
\catcode`\›=\active\def›{\~o}        
\catcode`\Ì=\active\defÌ{\~A}        
\catcode`\"=\active\def"{\~N}        
\catcode`\Í=\active\defÍ{\~O}        
\catcode`\‰=\active\def‰{\^a}        
\catcode`\=\active\def{\^e}        
\catcode`\"=\active\def"{\^{\i}}     
\catcode`\™=\active\def™{\^o}        
\catcode`\ž=\active\defž{\^u}        

\let\optionkeymacros\null

\begin{document} 
\begin{flushright} 
{OITS 610}\\
{January 1997}
\end{flushright}
\vspace*{1cm}

\begin{center} {\Large {\bf Fluctuations of Spatial Patterns as a
Measure of Classical Chaos}}
\vskip .75cm
 {\bf Zhen Cao and Rudolph C.\ Hwa }
\vskip.5cm
 
{Institute of Theoretical Science and Department of
Physics\\ University of Oregon, Eugene, OR 97403-5203,
USA}
\end{center}

\begin{abstract}
In problems where the temporal evolution of a nonlinear system
cannot be followed, a method for studying the fluctuations of
spatial patterns has been developed.  That method is applied to
well-known problems in deterministic chaos (the logistic map
and the Lorenz model) to check its effectiveness in characterizing
the dynamical behaviors.  It is found that the indices $\mu_q$
are as useful as the Lyapunov exponents in providing a
quantitative measure of chaos.

\end{abstract}

\section{Introduction} 

An important feature of classical nonlinear systems is that a
trajectory traced out by time evolution is well defined, so the
distance between nearby trajectories is a meaningful function of
time.  The Lyapunov exponents that characterize the distance
function have therefore been used widely to describe the chaotic
behaviors of such systems.  Certain quantum systems, however,
do not have such a feature.  In particular, self-coupled
quantum fields such as those in the $\phi^3$ theory do not have
evolutionary histories that can readily be described by
trajectories, since the number of degrees of freedom changes
with time.  In such problems alternative criteria for chaos beside
the use of Lyapunov exponents must be found.  A measure useful
in the study of QCD parton showers is a set of indices $\mu_q$
that characterize the nature of fluctuations of spatial patterns
\cite{ch}.  It is the purpose of this paper to apply that measure
to classical nonlinear systems and investigate its usefulness as
an alternative criterion for chaos.

In microscopic quantum systems it is often impossible to track
the time evolution of their states without disturbing the
systems.  Instead, it is the final state that can be measured with
high accuracy.  A prime example of problems of that type is the
inelastic collision of elementary particles at very high energy,
where many particles are created.  The momenta of all charged
particles in the final state can be determined precisely in
experiments.  Thus for each collisional event the momenta of
those particles constitute a spatial pattern in momentum space. 
From event to event those patterns change not only in the
magnitudes and directions of the momentum vectors, but also
in the total number of such vectors.  The challenge has been in
finding an efficient way of characterizing the fluctuation of
those patterns in experiments where millions of events are
measured.  Moreover, it has been of interest to find out whether
the notion of chaos has any meaning for such multiparticle
production processes.

In order to answer the latter question, i.\ e., the meaning of
chaos for self-reproducing nonclassical systems, it is necessary to
apply a chosen measure of fluctuations in such systems to some
classical problems for which the criteria for chaos are well
known.  The issue becomes the following:  if a classical chaotic
system exhibits certain familiar characteristics in its time
evolution, what can be said about the nature of the spatial
patterns associated with its trajectories?  In finding an answer to
this question we shall have accomplished in making two
beginnings:  on the one hand, we shall gain some insight on
whether the concept of chaos can be generalized to include
self-reproducing quantum systems, and on the other, a new
approach to the study of classical chaos will be opened up.  The
latter is an unexpected bonus that results from the attempts to
deal with the demands and concerns of a very different field of
physics.

In order to render this paper self-contained, a review of the
measure of fluctuations will be given (in Sec.\ 2) without the
particle physics in which it is originated.  The body of this paper
is the application of that measure to the logistic map and
the Lorenz attractor \cite{sc}.  We compare the dependences of
the Lyapunov exponents $\lambda$ on the control parameter $r$
with those of the indices $\mu_q$.  It is the close
correspondence between the two measures for both
deterministic systems that supports our view on the usefulness
of $\mu_q > 0$ as a criterion for chaos.

\section{Entropy Indices $\mu_q$}

Consider the problem of describing a system by making many
experimental measurements, each of which is called an event. 
An event consists of a spatial pattern in $d$-dimensional space. 
Let $F_q$ be a measure of that pattern to be described below.  
From event to event $F_q$ can fluctuate.  After ${\cal N}$
events, a large number, one has a distribution of $F_q$, which
we denote by $P(F_q)$, normalized to 1.  By taking the
normalized moments of $P(F_q)$, defined by 
\begin{eqnarray}
C_{p,q} = \left<F_q^p\right>/\left<F_q\right>^p \quad ,
\label{1}
\end{eqnarray}
we have a quantification of the fluctuations of the spatial
patterns.

Returning to the definition of $F_q$ itself, it is necessary to
recognize first that any description of a spatial pattern depends
on the resolution used. Let the $d$-dimensional space (call it
phase space, although it can be just the coordinate space, or just
the momentum space, or both) be divided into $M$ bins, each
having a volume $V_{\rm bin} = \delta^d$. 
Furthermore, let the intensity of the pattern be discretized at the
bin level so that at the $\it{i}$th bin the bin multiplicity
\begin{eqnarray}
n_i = \int_{V_i} \rho \left(\vec{r} \right) d^dr
\label{2}
\end{eqnarray}
is rounded out to an integer, where $\rho\left(\vec{r} \right)$ is
the density at the point $\vec{r}$.  For each event $F_q$ is
defined, for any integer $q \geq 2$, by 
\begin{eqnarray}
F_q = {1 \over M} \sum^M_{i=1} n_i \left( n_i - 1  \right) \cdots
\left( n_i - q + 1 \right)/ \left({1 \over M}\sum^M_{i=1}
n_i\right)^q
\label{3}
\end{eqnarray}
If $Q_n$ denotes the distribution of bin multiplicity $n$ in the
$M$ bins, normalized to $\sum_nQ_n = 1$, then $F_q$ can also be
written as
\begin{eqnarray}
F_q = \left< n^{[q]}\right>_h/ \left< n\right>^q_h
\label{4}
\end{eqnarray}
where $n^{[q]} = n!/(n-q)!$ and $\left< \cdots \right>_h$ is a
(horizontal) average over $Q_n$.  By horizontal, we mean
averaging over the multiplicity distribution in a given event, to
be distinguished from vertical averaging, such as in (\ref{1}),
which is an average over all events.

The virtue of the normalized factorial moments $F_q$ is that
they are trivial for statistical fluctuations \cite{bp}.  Let $Q_n$ be
a Poisson transform
\begin{eqnarray}
Q_n = \int^{\infty}_0 {s^n \over n!} \, e^{-s} D(s) ds \quad , 
\label{5}
\end{eqnarray}
where $D(s)$ may be regarded as some dynamical distribution,
whose convolution with the Poisson distribution (of statistical
origin) gives rise to the observed $Q_n$.  It is clear that, since 
\begin{eqnarray}
\left< n^{[q]}\right>_h = \int^{\infty}_0 s^q  D(s)
ds \quad , 
\label{6}
\end{eqnarray}
trivial dynamics represented by $D(s) = \delta (s-\bar{n})$
results in $F_q = 1$ for all $q$.  Indeed, (\ref{6}) indicates that
the statistical fluctuation is filtered out from the factorial
moments, yielding just the simple moments of the dynamical
$D(s)$.  Thus $F_q$ provides an effective description of the
dynamical fluctuations that generate the spatial pattern under
study.

Now let us consider the nature of the fluctuations from event to
event.  First, (\ref{1}) can be rewritten in the form 
\begin{eqnarray}
C_{p,q} = \left<\Phi_q^p\right>\quad, \qquad \Phi_q = 
{F_q \over \left<F_q\right>}
\quad .
\label{7}
\end{eqnarray}
While much information can be revealed by studying all
moments
$p$ of $P(F_q)$, it is sufficient for our purpose here to examine
only the neighborhood of $p = 1$.  It is analogous to studying the
information dimension $D_1$, which is the fractal dimension at
order 1 \cite{her}.  With the definition
\begin{eqnarray}
\Sigma _q =\left.{d \over dp} C_{p,q}\right|_{p=1} \quad ,
\label{8}
\end{eqnarray}
we have, on the one hand,
\begin{eqnarray}
\Sigma _q = \left<\Phi_q {\rm ln} \Phi_q\right> \quad .
\label{9}
\end{eqnarray}
On the other hand, if $C_{p,q}$ has a power-law behavior in $M$, 
\begin{eqnarray}
C_{p,q}  \propto M^{\psi_q(p)}\quad ,
\label{10}
\end{eqnarray}
which has been referred to as erraticity \cite{hwa}, then we also
have
\begin{eqnarray}
\Sigma _q \propto \left.{d \over dp} \psi_q(p)\right|_{p=1} {\rm
ln} M 
\quad .
\label{11}
\end{eqnarray}
For brevity we define
\begin{eqnarray}
\mu _q = \left.{d \over dp} \psi_q(p)\right|_{p=1} \quad , 
\label{12}
\end{eqnarray}
and refer to them as entropy indices.  It then follows that 
\begin{eqnarray}
\mu _q = {\partial \Sigma_q \over \partial {\rm ln} \, M}
\label{13}
\end{eqnarray}
in the scaling region, i.\ e., where $\Sigma_q$ exhibits a linear
dependence on ${\rm ln} M$.  It is not difficult to show how
$\mu _q$ is related to an entropy defined in the event space
\cite{ch,her}, but that connection is not needed here.

If there is no strict scaling behavior in $M$, then (\ref{10}) may
have to be generalized to accommodate a possible scaling law in
$g(M)$
\begin{eqnarray}
C_{p,q}  \propto g(M)^{\psi_q(p)}\quad ,
\label{14}
\end{eqnarray}
where $g(M)$ is some function of $M$.  In such cases $\Sigma_q$
and $\mu_q$ are defined as in (\ref{8}) and (\ref{12}), the only
difference being that $M$ is replaced by $g(M)$ in  (\ref{11})
and (\ref{13}).  Thus, instead of (\ref{11}), we would have 
\begin{eqnarray}
\Sigma _q (M, r)\propto \mu_q(r) {\rm ln} g(M) 
\quad ,
\label{15}
\end{eqnarray}
where we have introduced a control parameter $r$, the
dependence on which has been assumed implicitly in the
foregoing, but will become explicit in the following sections.  The
factorizable form of (\ref{15}) suggests that $g(M)$ may be
determined from $\Sigma _q (M, r)$ by evaluating it at a
particular $r_0$ so that 
\begin{eqnarray}
\Sigma _q (M, r) \propto \beta_q(r) \Sigma _q \left(M, r_0\right)
\label{16}
\end{eqnarray}
where
\begin{eqnarray}
\beta_q(r)= \mu_q(r) / \mu_q\left(r_0\right)
\label{17}
\end{eqnarray}
In this way $\mu_q (r)$ can be determined only up to an
overall factor for all $r$ .

We have described above a procedure by which one can take
${\cal N}$ events of fluctuating, spatial patterns, and by using 
(\ref{3}),  (\ref{7}), (\ref{9}) and (\ref{13}) [or (\ref{16}) and  
(\ref{17})] determine a set of indices $\mu_q$, $q = 2, 3, \cdots$,
that can efficiently characterize the nature of the fluctuations. 
In practice, it is not necessary to examine a large number of
$\mu _q$; $\mu_2$ and $\mu_3$ should suffice.  In the
following sections we shall use $\mu_2$ as a measure to study
the properties of the logistic and Lorenz problems and compare
its behaviors with those of the Lyapunov exponents $\lambda$.

\section{The Logistic Map} 

The simplest and well-understood example of deterministic
chaos is the logistic map \cite{sc,ce}.  We consider this example
to illustrate the use of $\mu_2$, since the value of $\lambda$
for it is well known and can therefore readily provide a
comparison with our result on $\mu_2$.

In the 1-dimensional interval $0 < x <1$, the map is 
\begin{eqnarray}
x_{j+1} = r\, x_j\, \left( 1 - x_j\right) \quad .
\label{18}
\end{eqnarray}
By repeated iteration one generates a sequence $T(x_0) =
\{x_0, x_1, \cdots, x_j, \cdots \}$, starting from a chosen initial
point $x_0$.  Every such sequence can be regarded as a
trajectory as time evolves, where the time is identified with the
number of iterations.  The distance $d_j$ between two
trajectories $T$ and $T^{\prime}$ is $\left| x_j - x_j^{\prime}
\right|$ at the $\it{j}$th step.  For $r > r_c = 3.5699456
\cdots$, but $< 4$, $d_j$ can grow exponentially for two nearby
trajectories with $d_0 = \left| x_0 - x_0^{\prime}\right| =
\epsilon$ infinitesimally small.  Except for certain narrow
intervals between $r_c$ and $4, \lambda$ is positive, and the
system exhibits chaotic behavior.

The first question to face is how such a behavior in time
evolution can be treated from the point of view of spatial
patterns, which is what $\mu_q$ are designed to describe.  Since
a trajectory in this case is automatically a collection $T(x_0)$ of
discrete points in $x$, the answer is, of course, obvious.  A
judicious choice of a subset of $T(x_0)$ is a spatial pattern of
interest, and each event corresponds to a particular initial value
$x_0$.  To see what subset is appropriate, we show in Fig. 1 a
plot of $d_j$ vs $j$ for $r = 3.99$ and for various small values
of $d_0$.  The value of $\lambda$ can be read off from the
initial exponential growth, $d_j = d_0 e^{j\lambda}$, to be
$\lambda = 0.66$, very close to the analytical value ${\rm ln} 2$
at
$r = 4$.  A significant aspect of Fig. 1 is that even for $d_0 = 10
^{-12}$ it takes only 40 time steps for $d_j$ to reach $O(1)$,
beyond which $d_j$ fluctuates with no apparent order.  At
smaller values of $r$, but above $r_c$, $\lambda$ would be
smaller and it takes longer for $d_j$ to get beyond the
exponential growth phase.  Two spatial patterns having
infinitesimal $d_0$ would be nearly the same if the
corresponding subsets of $T(x_0)$ and
$T^{\prime}(x_0^{\prime})$ consist of only the points in the
growth phase.  To exhibit chaotic behavior it is necessary that $j
> \lambda ^{-1} {\rm ln} d_0^{-1}$, so our subset $S({x_0})
\subset T(x_0)$ should consist of points above that value of $j$. 
Since we want to study the relationship between $\lambda$ and
$\mu_2$ for all interesting values of $r$, our choice of points for
$S({x_0})$ is as follows 
\begin{eqnarray}
S({x_0}) = \left\{ x_{\Delta}, x_{2\Delta},\cdots,
x_{m\Delta}\right\}_{x_0} \quad ,
\label{19}
\end{eqnarray}
where $\Delta = 100$, and $m = 20$.  Each event of that type has
a specific $x_0$, not included in $S({x_0})$.  We generate
${\cal N} = 10^5$ events whose initial $x_0$ are all randomly
generated within a small interval $(X_0, X_0 + 10^{-5})$ around
an arbitrarily chosen value $X_0$.  For the results to be shown
below, $X_0$ is $0.35436$.  Thus all ${\cal N}$ events
correspond to initially nearby trajectories, the distances between
any two of which diverge after a certain number of steps.

For each of the ${\cal N}$ events generated according to the
prescription described above, we divide the unit interval into
$M$ bins of $\delta$ size, count the number of points that fall
into each bin, and calculate $F_q(M)$ for that event by use of
(\ref{3}).  Then $\Sigma _q(M)$ is determined by performing
the appropriate vertical averaging in (\ref{9}).  With focus on
$q = 2$, the dependence of  $\Sigma _2(M)$ on ${\rm ln} M$ is
shown in Fig. 2(a) for a few representative values of $r$. 
Evidently, there is no linear dependence.  We thus use the
generalized scaling form expressed in (\ref{14}) and consider the
plot of (\ref{16}).  That is done in Fig. 2(b), which shows good
linear behavior.  The value of $r_0$ is chosen to be 3.9.  The
slopes $\beta _2 (r)$ can be determined from the best fits of all
the points for each $r$, and give, by (\ref{17}), the values of
$\mu_2(r)$ apart from a multiplicative constant.

Figure 3 shows the comparison of $\lambda$ and $\mu _2$,
where the overall normalization of $\mu _2$ in the figure is
adjusted to agree with $\lambda$ at $r = 3.8$.  The error bars on
the values of $\mu _2$ are due to the deviations from strict
straightlines in Fig. 2(b).  Clearly, $\mu_2(r)$ agrees very well
with $\lambda(r)$ throughout the whole range of $r$, except
that when $\lambda(r) \leq 0, \mu _2(r)$ can only be zero,
since it is a nonnegative quantity.

It is by virtue of Fig.\ 3 that we infer the effectiveness of using
the positivity of $\mu _2$ as a criterion for chaos.  In fact, $\mu
_q$ for higher $q$ have the same property, but they are not
needed for the simple system under consideration.  Thus we
conclude that the fluctuations of spatial patterns can be used to
reveal the chaotic behavior through the study of $\mu_q$ as
much as one can learn from the temporal evolution of nearby
trajectories.

\section{Lorenz Attractor}

We now consider another problem to explore the effectiveness of
$\mu _q$ in a dissipative dynamical system.  The prime
example of such systems is the Lorenz model, described by the
following equations:
\begin{eqnarray}
\dot{x} &=& -\sigma (x - y)  \quad , \nonumber \\ 
\dot{y} &=& rx - y - xz  \quad , \nonumber \\                     
\dot{z} &=& -bz + xy  \quad .
\label{20}
\end{eqnarray}
We fix, as with Lorenz \cite{en}, $\sigma = 10$ and $b = 8/3$,
and vary $r$ as the control parameter.  We discretize the time
variable and solve (\ref{20}) by repeated iterations starting
from some arbitrary point away from the fixed points.  The
critical value
$r_c$ of the control parameter, above which the trajectory
becomes unstable, depends on the size of the time step $\delta
t$ used.  It is found that $r_c$ increases slowly when $\delta t$ is
decreased. For computational efficiency we have chosen $\delta t
= 10^{-3}$.  Figure 4 shows how rapidly the $t$ dependence of
the distance function $d(t)$ changes, when $r$ is increased
infintesimally from below to above $r_c$.  We determine the
value of
$\lambda$ from straightline fits of the rising portions of log
$d(t)$ for every value of $r$ examined.  However, because log
$d(t)$ does not rise linearly with $t$ for $r > r_c$ a range of
values of $\lambda$ can be extracted from the fits.  We shall
indicate the result by shaded bands in $\lambda(r)$.

We use the same technique as described in Sec.\ 3 to generate a
spatial pattern for each event.  For $r > r_c$ the trajectory is the
familiar Lorenz attractor.  Since it is in 3D, we select 70 points
spaced 1 time unit apart (i.\ e.\ $10^3$ time steps of $\delta t$),
and then make a projection of them to the $x$-$y$ plane.  Figure
5 shows a typical event.  A total of $10^4$ events are generated,
each of which starts out initially at a random point in a small
cube of size $10^{-10}$ on each side, located at the point  $x_0 =
0$, $y_0 = 1$, $z_0 = 0$.  Since the Lorenz attractor is confined to
a finite region of space, which, when projected onto the $x$-$y$
plane, shows the points mainly along the diagonal of $x \approx
y$.  We have rotated the coordinates by $\pi/4$ so that the
pattern of points is mainly along the new $x$ axis shown in Fig.\
5 $(-30
\leq x \leq 30)$ with a dispersion in the expanded new $y$-axis 
$(-10 \leq y \leq 10)$.  This $2D$ rectangular space is divided
into $M$ square bins, and the multiplicity $n$ of points in each
bin is counted for the computation of $F_q$ in (\ref{3}) for each
event.  Using the procedure described in Sec. 2 the quantity
$\sum_2$ is determined and plotted against log $M$ in Fig. 6(a)
for various values of $r$.  Scaling is obtained by plotting against
$\sum_2(r_0)$, as in Fig. 6(b), where $r_0$ is chosen to be 28. 
From the slopes of the lines in the latter figure the indices
$\mu_2(r)$ are determined apart from an overall factor, which
is fixed by normalizing $\mu_2(r) = \lambda (r)$ at $r = 22.9$.

Figure 7 shows the results of our calculations of both
$\lambda(r)$ and $\mu_2(r)$.  As mentioned earlier, because of
the complicated $t$ dependence of $d(t)$, there is a band of
values of $\lambda$ for each $r$.  We have determined
$\lambda(r)$ only for some representative values of $r$.  Given
the errors involved, the agreement between $\lambda(r)$ and
$\mu_2(r)$ should be regarded as being quite good.  The most
important point is that they both show stepwise increase at
$r_c$.  Thus the utility of the positivity of $\mu_2(r)$ as a
criterion for chaos is clearly as effective as that of $\lambda(r)$.

\section{Large $M$ Behavior}

In the previous two examples we determine the slopes
$\beta_2(r)$ from Figs.\ 2 and 6 and by use of (\ref{16}); from 
$\beta_2(r)$ we obtain $\mu_2(r)$ apart from an overall
constant.  What we want to emphasize here is that the scaling
behaviors are for a range of $M$ that is not asymptotically large,
i.e., bin size $\delta$ is not infinitesimally small.  For generic
problems in statistical physics and fractal geometry, the
extension toward larger values of $M$ is the conventional
procedure.  However, for problems that we consider here such
an extension is inappropriate.  To explain that is the aim of this
section.

In fractal geometry, for example, one can take the mathematical
limit of smaller and smaller scale.  The fractal object can always
be examined with finer and finer resolution.  But in high-energy
physics, on the other hand, the number of particles produced in
any collisions is finite at finite energy.  In the limit $\delta
\rightarrow 0$ the bin multiplicities can only be $0$ and $1$,
and all $F_q = 0$ for $q \geq 2$.  For the logistic and Lorenz
problems that we have examined, we have taken a finite
number of points (20 and 70 respectively) to display the spatial
patterns.  Thus the $M \rightarrow \infty$ limit would also be
inappropriate.  Knowing exactly where all the points are in phase
space gives too much information and is not what we seek to
determine as the measure that can inform us about chaotic
behavior.

What can one say about the large $M$ regions above those
considered in Figs. 2 and 6, but not large enough to render all
$F_q = 0$?  We assert that they are of no dynamical interest. 
For $q = 2$ it is only necessary to examine the $M$ region in
which the bins are small enough to contain two or less points in
each bin, but not more.  Let $M^e_n$ be the number of bins in the
{\it e}th event with multiplicity $n$.  Then for that event we
have
\begin{eqnarray}
F_2 = {1 \over M} \sum_j n_j \left( n_j -1 \right)/\left(N \over
M\right)^2 = 2 MM^e_2/N^2 \quad ,
\label{21}
\end{eqnarray}
where $N$ is the total number of points in the event.  If ${\cal
N}_2$ denotes the number of events out of the total ${\cal N}$
events in which $M_2 \neq 0$, but $M_n = 0$ for $n \geq 3$,
then we get
\begin{eqnarray}
\left< F_2\right> = {2M \over {\cal N} N^2} \sum_e M^e_2 =
2Mr_2 \left< M_2\right>/N^2 \quad ,
\label{22}
\end{eqnarray}
where
\begin{eqnarray}
\left< M_2 \right> = {1 \over {\cal N}_2} \sum_{e\in {\cal
N}_2} M^e_2 \quad ,
\label{23}
\end{eqnarray}
and $r_2 = {\cal N}_2/{\cal N}$ is the fraction of events for
which $F^e_2 \neq 0$, but $F^e_{q>2} = 0$.  From (\ref{7}),
(\ref{21}) and (\ref{22}) we have 
\begin{eqnarray}
\Phi^e_2 = M^e_2/r_2\left< M_2 \right> \quad ,
\label{24}
\end{eqnarray}
so that from (\ref{9}) follows
\begin{eqnarray}
\Sigma_2 = {1 \over {\cal N}_2} \sum_e B_e {\rm ln} B_e - {\rm
ln} r_2 \quad ,
\label{25}
\end{eqnarray}
when $B_e = M^e_2/\left< M_2\right>$.  In the limit of large $M$
when $M^e_2 \rightarrow 1$ for nearly all events, then $B_e
\rightarrow 1$, and 
\begin{eqnarray}
\Sigma_2 \sim -{\rm ln}r_2 \quad .
\label{26}
\end{eqnarray}
Now, the probability for a bin in such events to have $n = 2$ is
$M^{-2}$.  Since it can be for any of the $M$ bins, we have 
\begin{eqnarray}
r_2 \sim M^{-1} \quad .
\label{27}
\end{eqnarray}
It then follows from (\ref{13}) that 
\begin{eqnarray}
\mu _2 = 1
\label{28}
\end{eqnarray}
The same line of reasoning leads also to the result
\begin{eqnarray}
\mu_q = q - 1 \quad .
\label{29}
\end{eqnarray}
In our numerical computation we have verified this result in
that Figs. 2(a) and 6(a) exhibit straightline behavior at large
$M$ with unit slope for all values of $r$.  Since only probabilistic
arguments have been used to derive the result, it is independent
of the structure of the model.

Thus in the search for scaling behavior in problems where $N$ is
finite one should not go to the extreme large $M$ region just
before all $F_q \rightarrow 0$, even though $\sum_2$ exhibits
linear dependence on ${\rm ln}M$ there.  The behavior that is
more relevant to the determination of $\mu_q$ involves $g(M)$,
defined in (\ref{14}), as the scaling variable, and it is the plots
like Figs. 2(b) and 6(b) that yield the more pertinent straightline
behaviors.

\section{Concluding Remarks}

By working with the two examples, the logistic map and the
Lorenz attractor, we have demonstrated that the index $\mu _2$
is as good as $\lambda$ in marking the chaotic regime of the
control parameters.  One may wonder why the complicated
procedure to determine $\mu _2$ should be considered when
the computation of $\lambda$ is significantly simpler.  We
reiterate that the rationale for studying spatial patterns is
rooted in the desire to examine chaotic behaviors in systems
where following the temporal evolution is not possible, or where
trajectories are ill defined. Such problems are far more
complicated than the simple nonlinear systems considered in
deterministic chaos.  The complexity of the procedure described
in Sec. 2 for the determination of the entropy indices $\mu _q$
is commensurate with the complexity of the problems.  Applying
such a tool to study the logistic map seems to be an overkill.  But
it has to be done in order to show the significance of the method. 
It is only when the agreement between $\mu _2$ and
$\lambda$ is established for problems with known behaviors of
$\lambda$ can one claim that $\mu _2 > 0$ is an effective
criterion for chaos, whether the system under study is simple or
complex.

Now that we have a method for treating the fluctuations of
spatial patterns, there seems to be a wide range of problems that
could not be studied effectively previously, but are now amenable
to analysis by this method.  They may range from cracks in dry
lake beds to galactic distribution.  When there is only one event,
like the astrophysical problem on galaxies, one should divide
the whole space into many subspaces, each constituting an event,
study the multiplicity fluctuations in bins of various sizes in
each subspace (event) and then average the fluctuations of those
patterns over all subspaces.  Even in problems of conventional
deterministic chaos, it is not always easy to fine-tune the initial
conditions experimentally.  Studying the properties of spatial
patterns may allow an experimentalist to circumvent the
fine-tuning difficulty.  It would be very interesting to explore
through the study of $\mu _2$ the possible universality among
many fields that have hitherto been regarded as totally
unrelated.

\section*{Acknowledgment} 

One of us (RCH) would like to thank T. Hwa for helpful
discussions.  This work was supported, in part,  by the U.\ S.\
Department of Energy under Grant No. DE-FG03-96ER40972.

\newpage
\begin{center}
\section*{Figure Caption}
\end{center}
\begin{description}

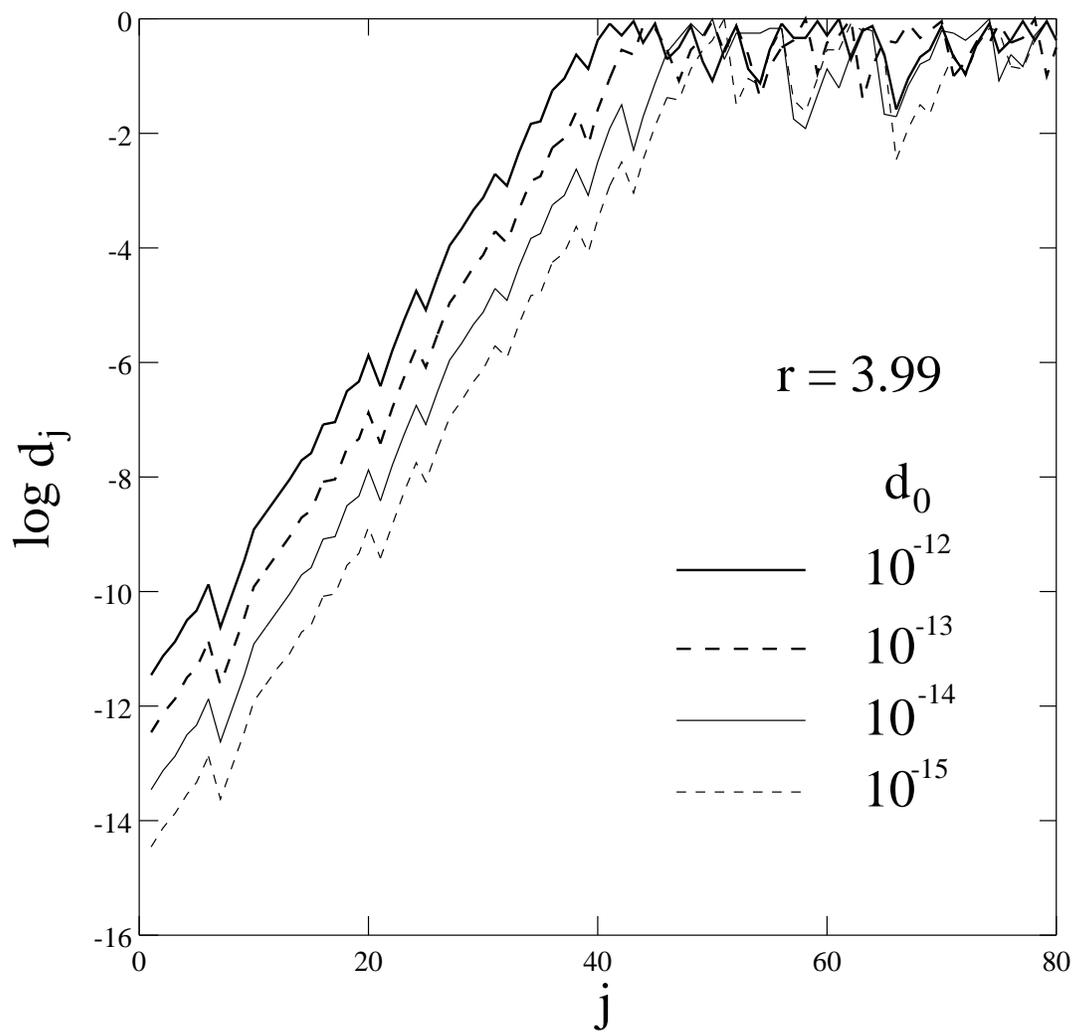
\item[Fig. 1]Exponential growth of the distance $d_j$ between
two trajectories as the time step $j$ is increased.

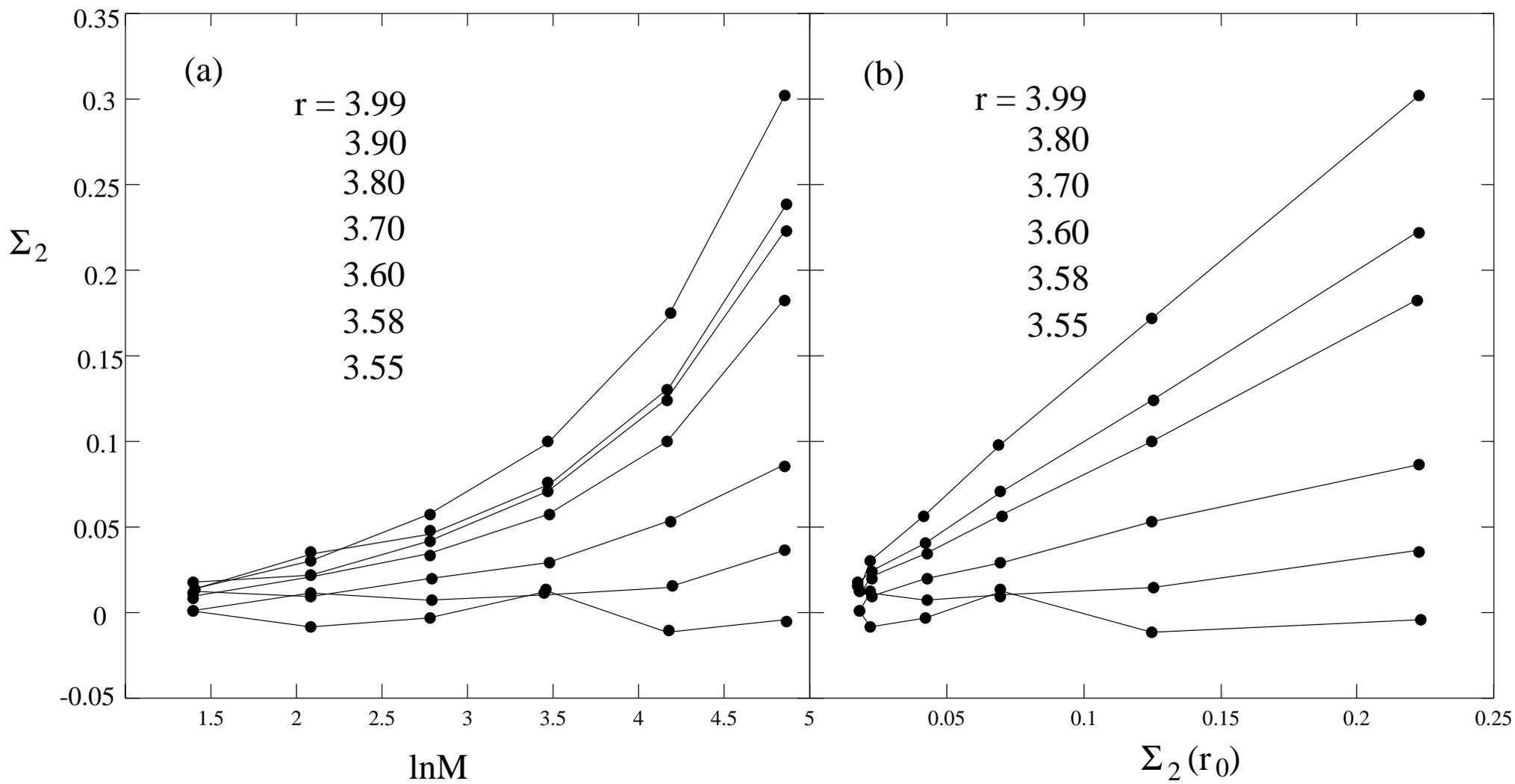
\item[Fig. 2]Behaviors of $\Sigma_2$ for the logistic map as a
function of (a)  ${\rm ln} M$ and (b)  $\Sigma_2\left(r_0\right)$
for various values of the control parameter $r$.  The value of
$r_0$ is chosen to be 3.9.

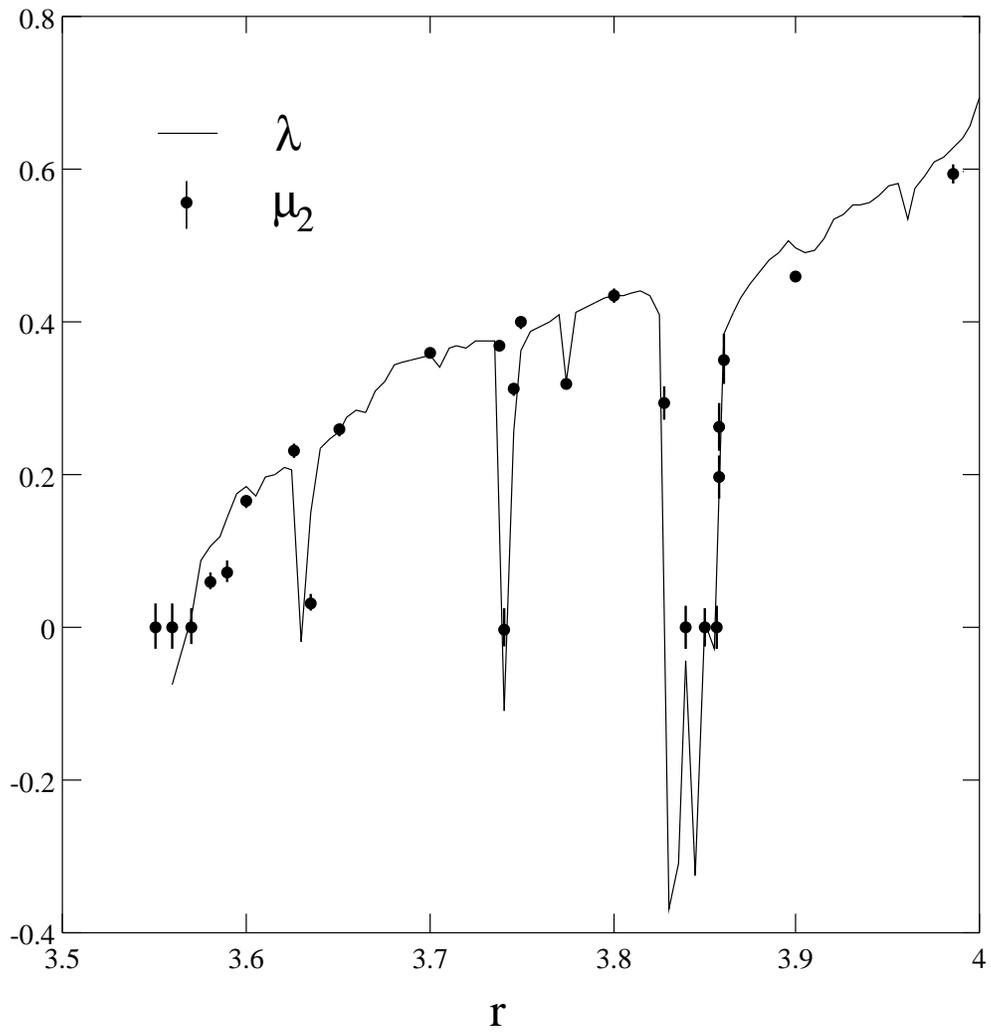
\item[Fig. 3]A comparison of $\mu _2$ with the Lyapunov
exponent $\lambda$ for the logistic map.

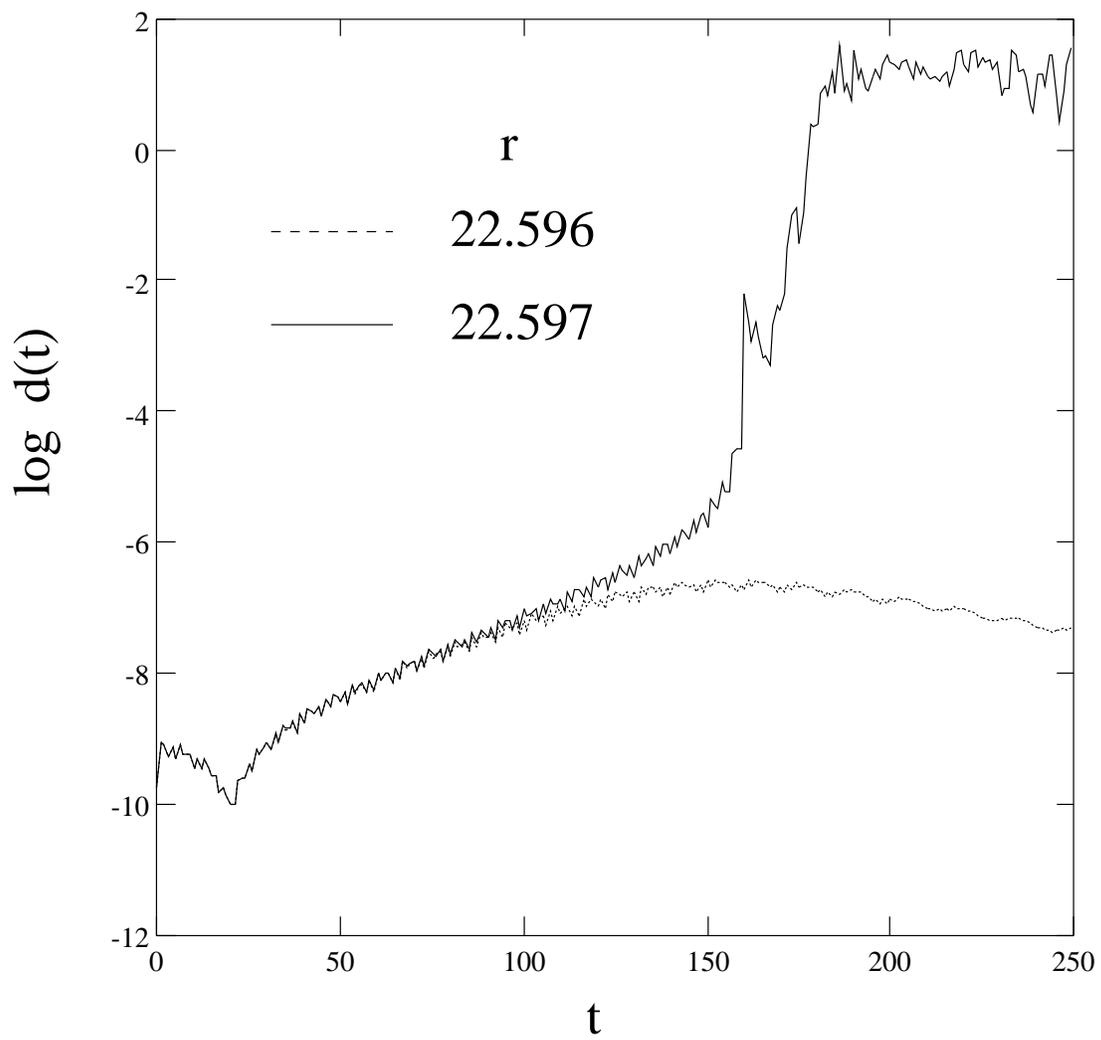
\item[Fig. 4]The behaviors of the distance function $d(t)$ for the
Lorenz attractor at two values of $r$ close to $r_c$.  

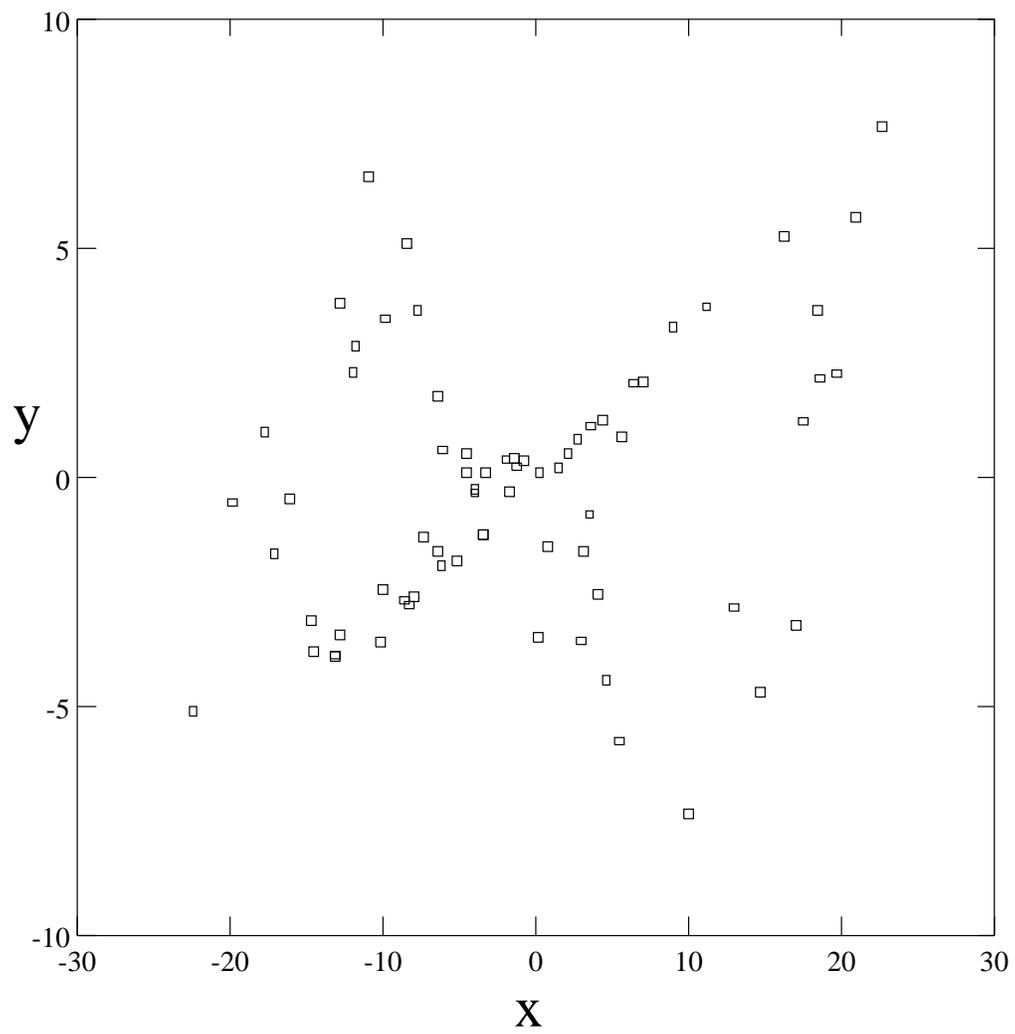
\item[Fig. 5]The spatial pattern of one event for the Lorenz
attractor when projected onto the $x$-$y$ plane and rotated by
$\pi/4$.

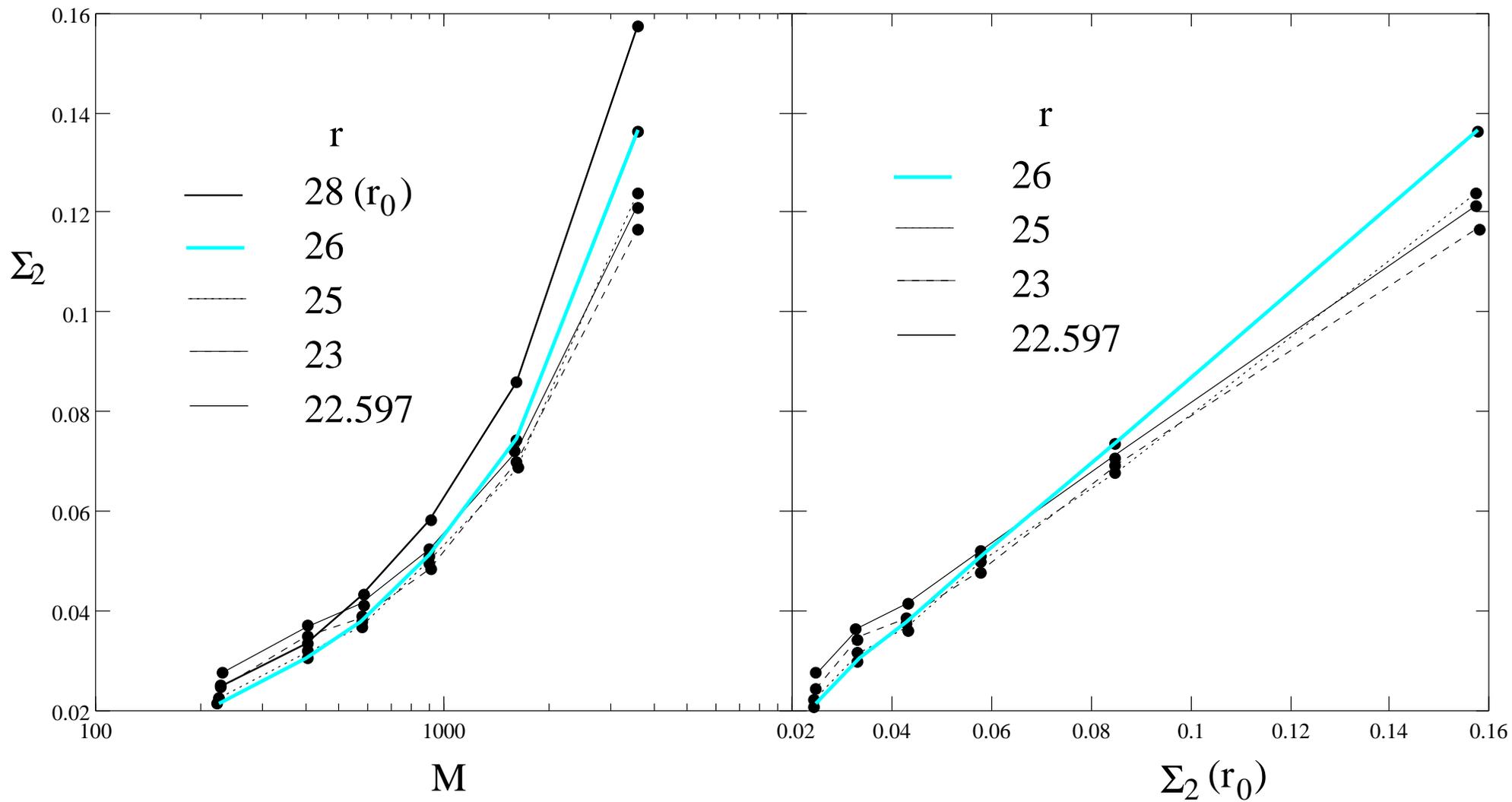
\item[Fig. 6]Same as for Fig. 2, but for the Lorenz attractor, and
with $r_0 = 28$.

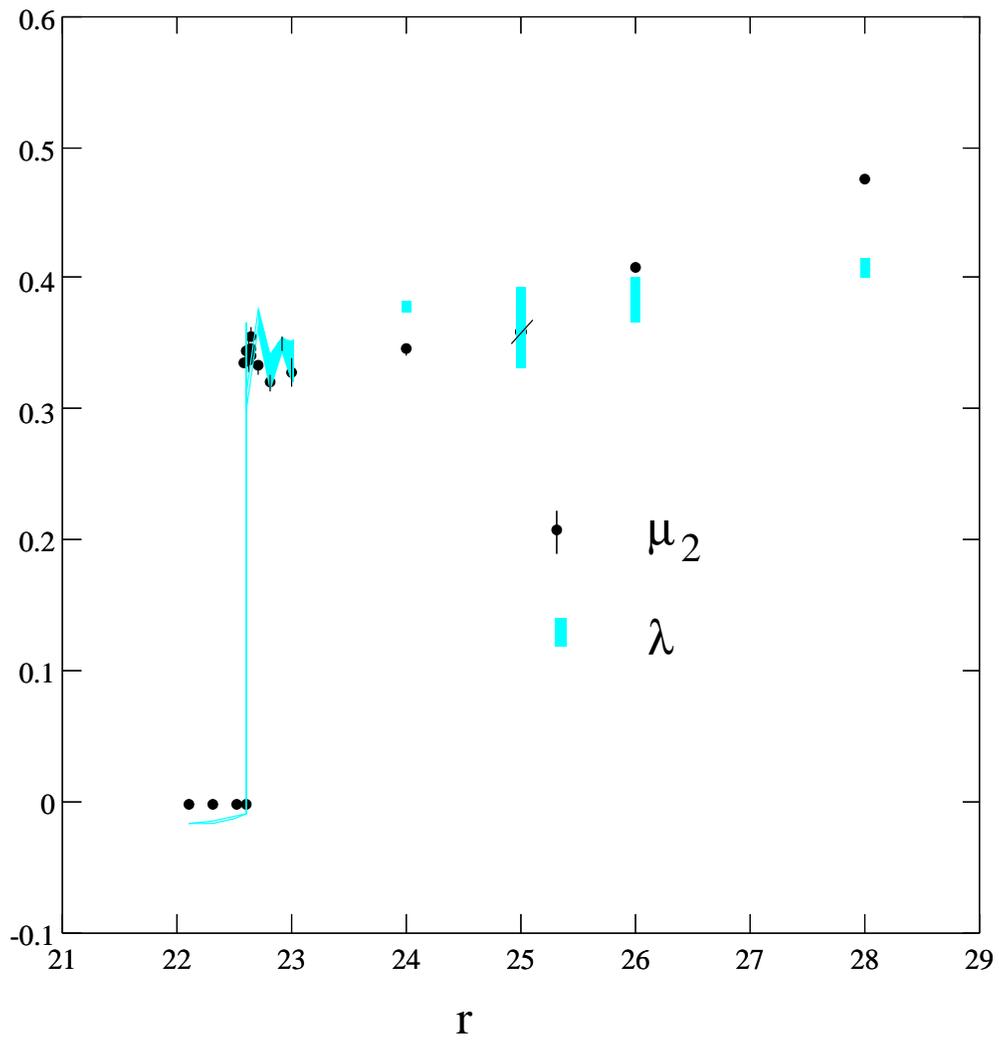
\item[Fig. 7]A comparison of $\mu _2$ with the Lyapunov
exponent $\lambda$ for the Lorenz attractor.

\end{description}


\begin{thebibliography}{99}


\bibitem{ch}Z.\ Cao and R.\ C.\ Hwa, Phy.\ Rev.\ Lett.\ {\bf 75},
1268, (1995); Phys.\ Rev.\ D {\bf 53}, 6608 (1996). 

\bibitem{sc}H.\ G.\ Schuster, {\it Deterministic Chaos}
(Physik-Verlag, Weinhein, 1984).

\bibitem{bp}A.\ Bia\l as and R.\ Peschanski, Nucl.\ Phys.\ {\bf
B273}, 703 (1986); {\bf B308}, 867 (1988).

\bibitem{her}H.\ G.\ E.\ Hertschel and I.\ Procaccia, Physica {\bf
8D}, 435 (1983).

\bibitem{hwa}R.\ C.\ Hwa, Acta Physica Polon.\ {\bf B27}, 1789
(1996).

\bibitem{ce}P.\ Collet and J.\ P.\ Eckman, {\it Iterated Maps on
the Interval as Dynamical Systems},
(Birkhauser, Boston, 1980).

\bibitem{en}E.\ N.\ Lorenz, J.\ Atmos.\ Sci.\ {\bf 20}, 130 (1963).

\end{thebibliography}
\end{document}